\documentclass[AMA,STIX1COL]{WileyNJD-v2}

\usepackage{moreverb}                 
\usepackage{url}  
\usepackage{float}
\usepackage{amssymb}
\usepackage{mathtools}
\usepackage{multirow}

\articletype{Research Article}

\received{}
\revised{}
\accepted{}

\begin{document}

\title{Imbalanced Randomization in Clinical Trials}

\author[1,2,3]{Thevaa Chandereng*}

\author[4]{Xiaodan Wei}

\author[1,2]{Rick Chappell}

\authormark{CHANDERENG \textsc{et al}}

\address[1]{\orgdiv{Department of Statistics}, \orgname{University of Wisconsin-Madison}, \orgaddress{\state{WI}, \country{USA}}}

\address[2]{\orgdiv{Department of Biostatistics \& Medical Informatics}, \orgname{University of Wisconsin-Madison}, \orgaddress{\state{WI}, \country{USA}}}

\address[3]{\orgdiv{Morgridge Institute of Research}, \orgaddress{\state{WI}, \country{USA}}}

\address[4]{\orgdiv{Biostatistics and Programming}, \orgname{Sanofi Bridgewater}, \orgaddress{\state{NJ}, \country{USA}}}

\corres{*Thevaa Chandereng, Department of Statistics, University of Wisconsin-Madison, Madison, WI, USA. \email{thevaasiinen@gmail.com}}

\abstract[Summary]{
Randomization is a common technique used in clinical trials to eliminate potential bias and confounders in a patient population. 
Equal allocation to treatment groups is the standard due to its optimal efficiency in many cases. 
However, in certain scenarios, unequal allocation can improve efficiency. 
In superiority trials with more than two groups, the optimal randomization is not always a balanced randomization. 
In non-inferiority trials, additive margin with equal variance is the only instance with balanced randomization. 
Optimal randomization for non-inferiority trials can be far from 1:1 and can greatly improve efficiency, a fact which is commonly overlooked. 
A tool for sample size calculation for non-inferiority trials with additive or multiplicative margin with normal, binomial or Poisson distribution is available at \href{http://www.statlab.wisc.edu/shiny/SSNI/}{http://www.statlab.wisc.edu/shiny/SSNI/}. 
}

\keywords{randomization, non-inferiority trial, efficiency}

\maketitle

\footnotetext{\textbf{Abbreviations:} NI, non-inferiority; ARE, asymptotic relative efficiency; CPORT, Cardiovascular Patient Outcomes Research Team; CI, confidence interval}

\section{Introduction}
Randomization remains a gold standard method in clinical trial methodology to eliminate potential bias and confounders.
Randomization eliminates a systematic difference between subjects in treatment groups inducing approximate balance with respect to covariates, both observed and unobserved.
Equal allocation to treatment groups is the standard due to its high efficiency in many cases. 
Statistical efficiencies are directly related to statistical power. 
Thus, increasing statistical efficiencies improves the likelihood of correctly rejecting the null hypothesis when the alternative is true.

However, the ethics of imbalanced randomization is highly debated. 
Unbalanced randomization is suggested to have an ethical advantage over balanced design due to the facts that more subjects are assigned to the new treatment than the control treatment and the new treatment is assumed to be superior \cite{pocock2013clinical}.    
However, there is a growing trend in trials with imbalanced treatment allocation \cite{hey2014questionable}.
Avins argue that if the chances of success higher with a new intervention, randomizing a greater proportion of subjects to the new intervention is highly desirable \cite{avins1998can}.
Pocock also asserts that there is a little loss in power with a moderate unbalanced randomization scheme \cite{avins1998can, pocock1979allocation}.
On the other hand, Edwards et al. argue that even though, unbalanced randomization might increase recruitment in a trial, the patient's expected benefit of a treatment might increase due to the notion of an increased chance of getting their preferred treatment \cite{edwards2000can}. 
Unbalanced randomization in a clinical trial might seem to be a favorable design due to several constraints such as limitation of available resources, increase chances of attaining required sample size and testing the side effects of a new treatment/drug. 

In certain trials, limitation of resources tends to hinder the success of a trial. 
Thus, investigators tend to reduce the patient allocation in the scarce group to overcome this issue. 
For instance, in the CPORT trial, if the availability of on-site cardiac surgery is a scarce resource, then the study could randomize more patients to the group of off-site cardiac surgery to overcome this issue. 
However, this was not the real reason behind the choice of 3:1 randomization in the CPORT study. 
Patients were randomized 3:1 to provide sufficient training to surgeons.
On the other hand, the cost can also contribute to the imbalance in a trial \cite{hey2014questionable}.  
Imbalanced allocation of patients might be beneficial to reduce the overall cost of a trial. 
Compared to inefficient trial, in an efficient trial, fewer patients are required to be recruited in a trial to attain the appropriate statistical power. 
 
Patients tend to be frustrated with equal allocation trials of a deadly disease due to the low chances of getting the new treatment \cite{hey2015outcome}. 
Thus, imbalanced trials tend to attract more patients due to a higher tendency of getting a new treatment. 
However, imbalance could reduce the statistical power and increase the sample size required \cite{hey2014questionable}.
Even though early phase trials are designed to study the efficacy of dosage of new treatment, these trials are conducted in a small sample and they fail to capture the complete effect of a treatment. 
Thus, imbalanced trials favoring an experimental treatment might be helpful to further analyze its side effects. 

After briefly looking at the optimal randomization ratio for multiple doses against control in superiority trials, we provide a quick introduction to non-inferiority trials. 
Then, we move on to the optimal allocation in a non-inferiority trials with equal and unequal variance and additive and multiplicative non-inferiority margins.  
We also derive the optimal randomization ratio for the number of events required for survival data in a non-inferiority trial before moving on to use the Cardiovascular Patient Outcomes Research Team (CPORT) study of percutaneous revascularization as an example. 

\section{"All-doses-against-control" in superiority trials}\label{sec1}
Dunnett's paper on comparing several treatments to a control illustrates the low efficiency of assigning all treatments equally for comparing multiple treatments to a control, but not to each other  \cite{dunnett1955multiple}. 
However, the paper did not provide a closed-form solution \cite{dunnett1955multiple}. 
This is derived here. 

\subsection{Equal variance}
Consider a superiority trial in which multiple experimental treatments (often different doses of the same drug) are compared to a control, with total sample size N and mean outcomes $\mu_i$, where $i$ indexes group, and constant variance $\sigma^2$.
There are a total of $k$ groups, where Group 1 is the control dose and the other $k-1$ groups are the experimental doses, and we want to compare the experimental groups to the control but not to each other.
The $k - 1$ null hypotheses are as follows, with alternatives of inequality. 
Our goal is to maximize their common power. 

\begin{eqnarray*}
H_{0,1}:\mu_2 &=& \mu_1, \\
H_{0,2}:\mu_3 &=& \mu_1, \\
&\vdots&         \\
H_{0,k-1}:\mu_k &=& \mu_1.
\end{eqnarray*}

Denote the sample size allocated in the control group $cN$, so that every other group is allocated $\frac{N(1 - c)}{k - 1}$. In order to compute the optimal $c$, we minimize the total variance below with respect to $c$ ($0 < c < 1$).

\begin{eqnarray*}
\sum_{i = 2}^{k} Var(\bar{X_i} - \bar{X_1})
&\propto& \frac{k - 1}{1-c} + \frac{1}{c} \\
\frac{\partial \sum_{i = 2}^{k} Var(\bar{X_i} - \bar{X_1})}{\partial c} &\propto& \frac{-1}{c^2} + \frac{k - 1}{(1- c)^2} = 0.
\end{eqnarray*}

\[c =
  \begin{dcases*}
   \frac{1}{2}, & if  k = 2\\
   \frac{-1 + \sqrt{k - 1}}{k - 2}, & if k > 2.
  \end{dcases*}
\]

The asymptotic relative efficiency (ARE) of the optimal allocation (A1) relative to equal allocation (A2) is

$$ARE(A1, A2) = \frac{2k(k - 1)}{(\sqrt{k - 1} + k - 1)^2} > 1, \quad \text{for }  k > 2.$$
 
 \subsection{Unequal variance}

If each of the dosage groups has a different variance, this becomes a harder optimization problem. Denote the variance of group $i$ by $\sigma_i^2$.  
The sample size allocated to each group is $c_i N$ for  group $i$ where $1 \leq i  \leq k$ ($0 < c_i < 1$, $\sum_{i = 1}^k c_i = 1$). 
The objective function is

\begin{eqnarray*}
\sum_{i = 2}^k Var(\bar{X_i} - \bar{X_1})
&\propto& \frac{(k - 1) \sigma_1^2}{c_1} + \frac{\sigma_2^2}{c_2} +...+ \frac{\sigma_k^2}{c_k}.
\end{eqnarray*}

The optimal solution for the $c_1$, ..., $c_{k - 1}$ is
\begin{eqnarray*}
c_1 = \frac{\sigma_1 \sqrt{k - 1}}{\sigma_1 \sqrt{k - 1} + \sum_{i = 2}^k \sigma_i},  \qquad  c_h = \frac{\sigma_h}{\sigma_1 \sqrt{k - 1} + \sum_{i = 2}^k \sigma_i} \qquad \text{for $h$ $\geq$ 2.}
\end{eqnarray*}

The asymptotic relative efficiency of the optimal allocation (A1) relative to equal allocation (A2) is

$$ARE(A1, A2) = \frac{k(k - 1) \sigma_1^2 + k \sum_{i = 2}^k \sigma_i^2}{(\sigma_1 \sqrt{k - 1} + \sum_{i = 2}^k \sigma_i)^2} > 1  \quad \text{for }  k > 2.$$

\section{Efficiency vs. randomization ratio in non-inferiority trials with two treatments}
\subsection{Non-inferiority trials}

Non-inferiority trials are clinical trials designed to establish that a new treatment is not that much worse than a standard control \cite{cook2007introduction}. 
Thus, a new treatment can be favorable even if it is slightly worse than the current treatment. 
Unlike equivalence trials, non-inferiority trials do allow the possibility of the new treatment being better than the standard treatment \cite{greene2008noninferiority}. 
The new treatment is believed to offer ancillary benefit in terms of side effects, safety or other factors. 
Outcomes are not always efficacy. 

A margin is introduced to allow a small loss in effect by the new treatment compared to the control treatment.
This margin serves as a maximum acceptable threshold and determined in advance. \cite{McDaniel2016, blackwelder1982proving}. 
Factors such as historical data and physician's experience play a vital role in the decision of setting a margin for a non-inferiority trial. 
The selection of margin scale used (difference in means of two groups, ratios of means, etc.) is less commonly stated but is important for computing sample sizes and interpretation of results \cite{chappell2012non}.

Denote $\Delta >  0$ for an additive margin and $\Delta > 1$ for a multiplicative margin. 
The corresponding non-inferiority trial hypotheses for an additive margin to test the difference in means are
\begin{eqnarray*}
H_0:\mu_C - \mu_T & \geq & \Delta \\
H_A:\mu_C - \mu_T & < \Delta
\end{eqnarray*}
where $\mu_C$ is the mean of the control group and $\mu_T$ is the mean of the experimental group. 
On the other hand, the corresponding non-inferiority trial hypotheses for a multiplicative margin to test the differences in means are 
\begin{eqnarray*}
H_0:\frac{ \mu_C}{\mu_T} & \geq  & \Delta \\
H_A:\frac{ \mu_C}{\mu_T} & < & \Delta
\end{eqnarray*}
where $\mu_C$ is the mean of the control group and $\mu_T$ is the mean of the experimental group. 
Higher means are considered favorable.

\subsection{Imbalance in non-inferiority trials: equal variance}

The optimal allocation for different scenarios are computed below.
The variance for the both the treatment and control groups is denoted by $\sigma^2$.

\subsubsection{Optimal allocation for additive $\Delta$}
\begin{eqnarray*}
H_0:\mu_C - \mu_T \geq \Delta.
\end{eqnarray*}

In this case, the optimal randomization ratio is 1:1 since $Var(\bar{X_C}) = Var(\bar{X_T})$ .

\subsubsection{Optimal allocation for multiplicative $\Delta$}
\begin{eqnarray*}
H_0:\frac{\mu_C}{\mu_T} \geq \Delta.
\end{eqnarray*}

The optimal sample size allocated in the control group is denoted $hN$ and every other group is allocated (1-$h)N$. In order to compute the optimal $h$, we minimize the variance below with respect to $h$, ($0 < h < 1$).

\begin{eqnarray*}
Var(\bar{X_T} - \Delta\bar{X_C}) &\propto& \frac{1}{h} + \Delta^2\frac{1}{1 - h} \\
\frac{\partial Var(\bar{X_T} - \Delta\bar{X_C})}{\partial h} &\propto& \frac{-1}{h^2} + \Delta^2\frac{1}{(1- h)^2} = 0  \\
h &=& \frac{1}{\Delta + 1}.
\end{eqnarray*}

The asymptotic relative efficiency of the optimal allocation (A1) relative to equal allocation (A2) is

\begin{eqnarray*}
ARE(A_1, A_2) &=& \frac{2(1 + \Delta^2)}{(\Delta + 1)^2} \geq 1.
\end{eqnarray*}

\subsection{Imbalance in non-inferiority trial: unequal variance}
The variance in the treatment groups are not equal, for example with binomial or Poisson outcomes. 
The variance is denoted by $\sigma_i^2 = V(\mu_i)$, $V(\mu_i)$ is the variance of treatment $i$ and it is a function of $\mu_i$.
\subsubsection{Optimal allocation for additive $\Delta$}
\begin{eqnarray*}
H_0:\mu_C - \mu_T \geq \Delta.
\end{eqnarray*}

The optimal sample size allocated in the control group is denoted $hN$ and every other group is allocated (1-$h)N$. In order to compute the optimal $h$, we minimize the variance below with respect to $h$, ($0 < h < 1$).

\begin{eqnarray*}
Var(\bar{X_C} - \bar{X_T}- \Delta) &\propto& \frac{\sigma_C^2}{h} + \frac{\sigma_T^2}{1 - h} \\
\frac{\partial Var(\bar{X_C} - \bar{X_T}- \Delta)}{\partial h} &\propto& \frac{-\sigma_C^2}{h^2} + \frac{\sigma_T^2}{(1 - h)^2} = 0  \\
h &=& \frac{\sigma_C}{\sigma_C + \sigma_T}.
\end{eqnarray*}

The asymptotic relative efficiency of the optimal allocation (A1) relative to equal allocation (A2) is

\begin{eqnarray*}
ARE(A_1, A_2)
&=& \frac{2(\sigma_C^2 + \sigma_T^2)}{(\sigma_C + \sigma_T)^2} \geq 1.
\end{eqnarray*}

\subsubsection{Optimal allocation for multiplicative $\Delta$}
\begin{eqnarray*}
H_0:\frac{\mu_C}{\mu_T} \geq \Delta.
\end{eqnarray*}

The optimal sample size allocated in the control group is denoted $hN$ and every other group is allocated (1-$h)N$. In order to compute the optimal $h$, we minimize the variance below with respect to $h$, ($0 < h < 1$).

\begin{eqnarray*}
Var(\bar{X_C} - \Delta\bar{X_T}) &\propto& \frac{\sigma_C^2}{h} + \Delta^2\frac{\sigma_T^2}{1 - h}\\
\frac{\partial Var(\bar{X_C} - \Delta\bar{X_T})}{\partial h} &\propto& \frac{-\sigma_C^2}{h^2} + \Delta^2\frac{\sigma_T^2}{(1- h)^2} = 0  \\
h &=& \frac{\sigma_C}{\sigma_C + \Delta \sigma_T}.
\end{eqnarray*}

The asymptotic relative efficiency of the optimal allocation (A1) relative to equal allocation (A2) is

\begin{eqnarray*}
ARE(A_1, A_2)
&=& \frac{2(\sigma_C^2 + \Delta^2\sigma_T^2)}{(\sigma_C + \Delta\sigma_T)^2} \geq 1.
\end{eqnarray*}

\subsection{The general case for generalized linear models}
The derivation for binomial and Poisson null hypothesis are illustrated below. 
In the binomial case, consider $\pi_C$ and $\pi_T$ as the probability of success for control and treatment group. 
Meanwhile, in the Poisson group consider $\lambda_C$ and $\lambda_T$ as the mean of control and treatment group respectively.

The optimal sample size allocated in the control group is denoted $hN$ and every other group is allocated (1-$h)N$. In order to compute the optimal $h$, we minimize the variance below with respect to $h$, ($0 < h 
< 1$).
\subsubsection{Additive hypothesis}

\begin{minipage}[t]{0.5\textwidth}

\centering{\textbf{\large Binomial}} 
\begin{eqnarray*}
H_0:  \pi_C - \pi_T \geq \Delta.
\end{eqnarray*}

\begin{eqnarray*}
h &=& \frac{\sqrt{\pi_C (1- \pi_C)} }{\sqrt{\pi_C (1- \pi_C)} + \sqrt{\pi_T (1- \pi_T)}}.
\end{eqnarray*}

The asymptotic relative efficiency of the optimal  \\ allocation (A1) relative to equal allocation (A2) is

\begin{eqnarray*}
ARE(A_1, A_2)
&=& \frac{2(\pi_C (1-\pi_C) + \pi_T(1-  \pi_T))}{(\sqrt{\pi_c(1-\pi_C)} + \sqrt{\pi_T (1- \pi_T}))^2} \geq 1.
\end{eqnarray*}

\end{minipage}
\begin{minipage}[t]{0.5\textwidth}

\centering{\textbf{\large Poisson}} 
\begin{eqnarray*}
H_0:  \lambda_C - \lambda_T \geq \Delta.
\end{eqnarray*}

\begin{eqnarray*}
h &=& \frac{\sqrt{\lambda_C} }{\sqrt{\lambda_C} +  \sqrt{\lambda_T}}.
\end{eqnarray*}

The asymptotic relative efficiency of the optimal allocation  \\(A1) relative to equal allocation (A2) is

\begin{eqnarray*}
ARE(A_1, A_2)
&=& \frac{2(\lambda_C +  \lambda_T)}{(\sqrt{\lambda_C} + \sqrt{\lambda_T})^2} \geq 1. \\
\end{eqnarray*}

\end{minipage}

\subsubsection{Multiplicative hypothesis}

\begin{minipage}[t]{0.5\textwidth}

\centering{\textbf{\large Binomial}} 

\begin{eqnarray*}
H_0: \frac{\pi_C}{ \pi_T} \geq \Delta.
\end{eqnarray*}

\begin{eqnarray*}
h &=& \frac{\sqrt{\pi_C (1- \pi_C)} }{\sqrt{\pi_C (1- \pi_C)} + \Delta \sqrt{\pi_T (1- \pi_T)}}.
\end{eqnarray*}

The asymptotic relative efficiency of the optimal allocation \\(A1) relative to equal allocation (A2) is

\begin{eqnarray*}
ARE(A_1, A_2)
&=& \frac{2(\pi_C (1-\pi_C) + \Delta^2 \pi_T(1- \pi_T))}{(\sqrt{\pi_c (1 - \pi_C)} + \Delta \sqrt{\pi_T (1- \pi_T}))^2} \geq 1.
\end{eqnarray*}

\end{minipage}
\begin{minipage}[t]{0.5\textwidth}

\centering{\textbf{\large Poisson}} 

\begin{eqnarray*}
H_0:  \frac{\lambda_C}{\lambda_T} \geq \Delta .
\end{eqnarray*}

\begin{eqnarray*}
h &=& \frac{\sqrt{\lambda_C} }{\sqrt{\lambda_C} + \Delta \sqrt{\lambda_T}}.
\end{eqnarray*}

The asymptotic relative efficiency of the optimal allocation (A1)\\ relative to equal allocation (A2) is

\begin{eqnarray*}
ARE(A_1, A_2)
&=& \frac{2(\lambda_C + \Delta^2 \lambda_T)}{(\sqrt{\lambda_C} + \Delta \sqrt{\lambda_T})^2} \geq 1.
\end{eqnarray*}

\end{minipage}

\subsection{Survival analysis}
In clinical trials of outcomes which are times to events and therefore subject to events and therefore subject to censoring sample size and efficiency depend on the number of events observed in each group.
We adopted the number of events required in the section below for two sample log-rank non-inferiority trials using proportional hazards assumption \cite{jung2005sample}. 
The hazard function for the control treatment is denoted by $\lambda_C$ and the hazard function for the treatment group is denoted by $\lambda_T$. 
Under the proportional hazards assumption, $ \Delta = \frac{\lambda_T (t)}{\lambda_C (t)}$ denotes the hazard ratio and usually the non-inferiority margin is set to $\Delta > 1$.

We obtained the derivation of the number of events required, D from Jung et al. (2005) \cite{jung2005sample}.
The test statistics is denoted by 
$$ H_0: \Delta \geq \Delta_0 \qquad\qquad H_A : \Delta < \Delta_0 .$$
Denote, $pN$ is the sample size allocated in the control group and $(1-p)N$ the sample size allocated in the treatment group. 
$\alpha$ and $1 - \beta$ corresponds to the respective type I error rate and power for the clinical study.

$$D = \frac{\{\sqrt{\Delta_0} z_{1 - \alpha} + (p + (1- p) \Delta_0) z_{1 - \beta})\}^2}{p(1-p)(\Delta_0 - 1)^2} $$

By substituting $a = \sqrt{\Delta_0} z_{1 - \alpha}$, $b = \Delta_0$, $c = z_{1 - \beta}$ and $d = (\Delta_0 - 1)^2$, we get

$$ D = \frac{\{a + (p + (1- p)b) c)\}^2}{p(1-p)d};  \qquad 0 > p > 1$$
\begin{eqnarray*}
\frac{\partial D}{\partial p} &\propto& 2(a + bc - bcp + cp)(c-bc)(1-p)pd - (a + bc -bcp+ pc)^2 ((1- 2p)d) \\
&\propto& 2c(a + bc - bcp + cp)(1-b)(1-p)p - (a + bc -bcp+ pc)^2 (1- 2p) \\
&=& (a + bc - bcp + cp) (2cp(1-b)(1-p) - (a + bc -bcp+ pc)(1- 2p)) \\
&=& (a + bc - bcp + cp)(cp - a - bc +bcp+ 2ap) . 
\end{eqnarray*}

Setting $\frac{\partial D}{\partial p}$ = 0 to obtain the minimum and replacing a, b and c respectively, we get 

\[p =
  \begin{dcases*}
   \frac{\sqrt{\Delta_0} z_{1 - \alpha} +  \Delta_0 z_{1 - \beta}}{( \Delta_0 + 1)z_{1 - \beta}  + 2\sqrt{\Delta_0} z_{1 - \alpha}} \\
   \frac{\sqrt{\Delta_0} z_{1 - \alpha} +  \Delta_0 z_{1 - \beta}}{ z_{1 - \beta} (\Delta_0 - 1)}.
  \end{dcases*}
\]

Since $0< p < 1$, the optimal solution for $p$, $p = \frac{\sqrt{\Delta_0} z_{1 - \alpha} +  \Delta_0 z_{1 - \beta}}{( \Delta_0 + 1)z_{1 - \beta}  + 2\sqrt{\Delta_0} z_{1 - \alpha}}$. 
The other solution of $p$ provides a value of greater than 1 when $\Delta_ 0$ > 1. 
The only instance with balanced randomization is when $\Delta_0 = 1$. However, this would not usually occur in a non-inferiority trial design. 
Chow et al. (2017) provides a derivation for number of events required where the assumuption of $S_C(t) \approx S_T(t)$ is made \cite{chow2017sample}. 
Due to the assumption, the optimal solution is $p$, $p = 0.5$ always as shown in the derivation below. 

$$ D = \frac{(z_{1- \alpha} + z_{1 - \beta})^2}{p(1-p) (log \Delta_0)^2} $$

\begin{eqnarray*}
\frac{\partial D}{\partial p} &\propto& \frac{- ((1 - p) (log \Delta_0)^2 - p (log \Delta_0)^2)}{(p(1-p) (log \Delta_0)^2)^2} \\ 
&=& \frac{(2p - 1) (log \Delta_0)^2}{(p(1-p) (log \Delta_0)^2)^2}. 
\end{eqnarray*}

Setting the numerator of $\frac{\partial D}{\partial p}$ = 0, we always get $p = 0.5$.

\section{Example: CPORT}
The Cardiovascular Patient Outcomes Research Team (CPORT) study provides a motivating example for the current work. 
In the CPORT study, physicians were interested in comparing the performance of percutaneous coronary intervention (PCI) at hospitals with vs. without on-site cardiac surgery \cite{Aversano2013}. 
PCI is often restricted to hospitals with on-site cardiac surgery which limits patients availability to receive the treatment.
Based on previous studies, the six weeks (all-cause) rate of mortality was estimated to be 0.8\%.
Therefore, the CPORT research tem was interested to show that $\pi_T - \pi_C <  0.004$

CPORT was a multi-center randomized trial with 18,867 patients who were randomized at 3:1 in the control group to undergo PCI with 14,149 patients undergoing PCI at a hospital without on-site cardiac surgery and 4,718 patients undergoing PCI at hospitals with on-site cardiac surgery. 
The six week mortality rate observed was 0.9\% at hospitals without on-site surgery and 1\% vice versa. 
The 95\% CI for the difference in six-week mortality rate was -0.31 to 0.23 with a p-value of 0.004 for non-inferiority.
The results suggest that PCI performed at hospitals without on-site cardiac surgery was non-inferior
to PCI performed at hospitals with on-site cardiac surgery with respect to mortality at six weeks and major adverse cardiac events at nine months.

\begin{figure}
\centering
\includegraphics[height=20pc,width=100mm]{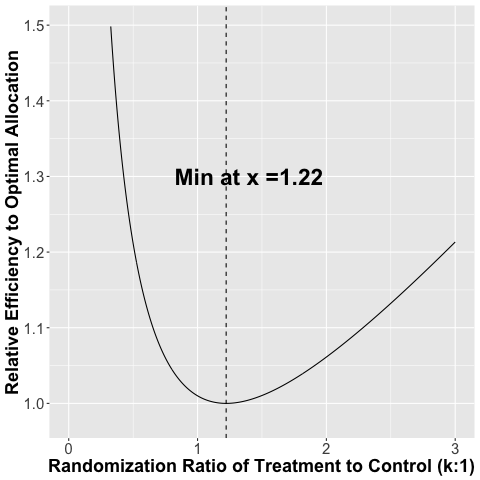}
\caption{Relative efficiency plot of CPORT trial versus randomization ratio of performing PCI at hospitals with no on-site cardiac surgery. The optimal randomization ratio is 1.22:1. }
\label{Fig:eff}
\end{figure}

However, the 3:1 randomization was not optimized statistically.
The optimal randomization ratio is 1.22:1 which is illustrated in Figure \ref{Fig:eff} using relative efficiency curve.
The computation of optimal allocation for different randomization ratio is illustrtaed in Figure \ref{Fig:eff} and Table \ref{tab:a}.

\begin{table}[htb]
    \centering
        \begin{tabular}{|c | c|}\hline
        Randomization Ratio of Treatment to Control (k:1) & Relative Efficiency to Optimal Allocation  \\
        \hline
        0.33 &  1.48 \\
        1.00 & 1.01 \\
        1.22 &  1.00\\
        3.00 &  1.21\\ \hline
        \end{tabular}
    \caption{Relative efficiencies of different randomization ratio to control (k:1). The loss of efficiency is about 1\% with equal allocation, however with 3:1 randomization ratio, the loss of efficiency increases to 21\%. Aversano et al. would have lost more efficiency if they decided to use 1:3 randomization ratio. }\label{tab:a}
\end{table}

\section{Discussion}

We have derived optimal allocation for superiority trials with multiple drugs versus control and computed the relative efficiency compared to a balanced allocation. 
The optimal allocation is unbalanced in a superiority trial with multiple drugs each compared to control. 
Even with superiority trials comparing two binomial endpoints, the optimal randomization is not 1:1 because the power calculations use the variance of the difference in proportions under the alternative, although this difference is usually trivial \cite{rosenberger2015randomization}. 
In non-inferiority trials, additive margin with equal variance is the only instance with balanced randomization.
With multiplicative margins, 1:1 randomization is less effective because we are estimating a weighted sum and even with additive margins in non-inferiority trials, variance differs under $H_0 $ for example with binary outcomes (binomial) implying optimality with unbalanced randomization.
In survival data, the balanced allocation is only optimal when the hazard ratio is equal to 1 which is unlikely in non-inferiority trials. 

As we demonstrated in the earlier sections, it is sometimes optimal to implement an unequal allocation. 
In designing clinical trials, the ethical concerns intertwine with proper scientific judgment.
Researchers have to be careful and wise in planning a trial to prevent failure in achieving the primary outcome. 

An application for a sample size calculator in a non-inferiority trial with randomization ratio and relative efficiency to balanced randomization is available at \href{http://www.statlab.wisc.edu/shiny/SSNI/}{http://www.statlab.wisc.edu/shiny/SSNI/}. 
The sample size calculator computes sample sizes for multiplicative or additive margin with normal, binomial or Poisson distribution. 
The application also reveals other randomization ratios with their relative efficiency compared to the optimal allocation. 

\begin{figure}
\centering
\includegraphics[height=15pc,width=140mm]{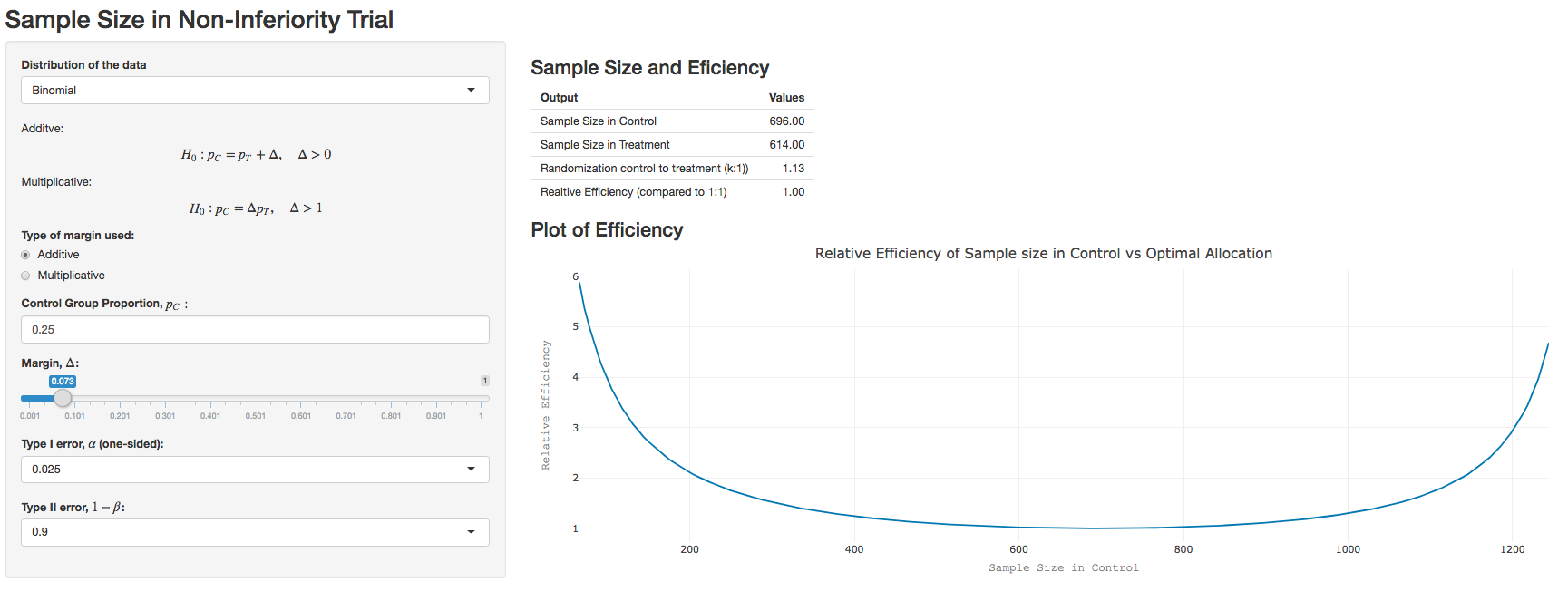}
\caption{The tool for sample size calculation for non-inferiority trial with additive or multiplicative margin with normal, binomial or poisson distribution.}
\label{Fig:ssni}
\end{figure}

\section*{Acknowledgments}
We are grateful to Lee McDaniel, Tom Cook, Marta Studnicka, Angela Zito, and Anthony Gitter for their helpful feedback and discussions.

\subsection*{Financial disclosure}

None reported.

\subsection*{Conflict of interest}

The authors declare no potential conflict of interests.

\section*{Supporting information}

The following supporting information is available as part of the online article:

\begin{figure}
\centering
\includegraphics[height=20pc,width=100mm]{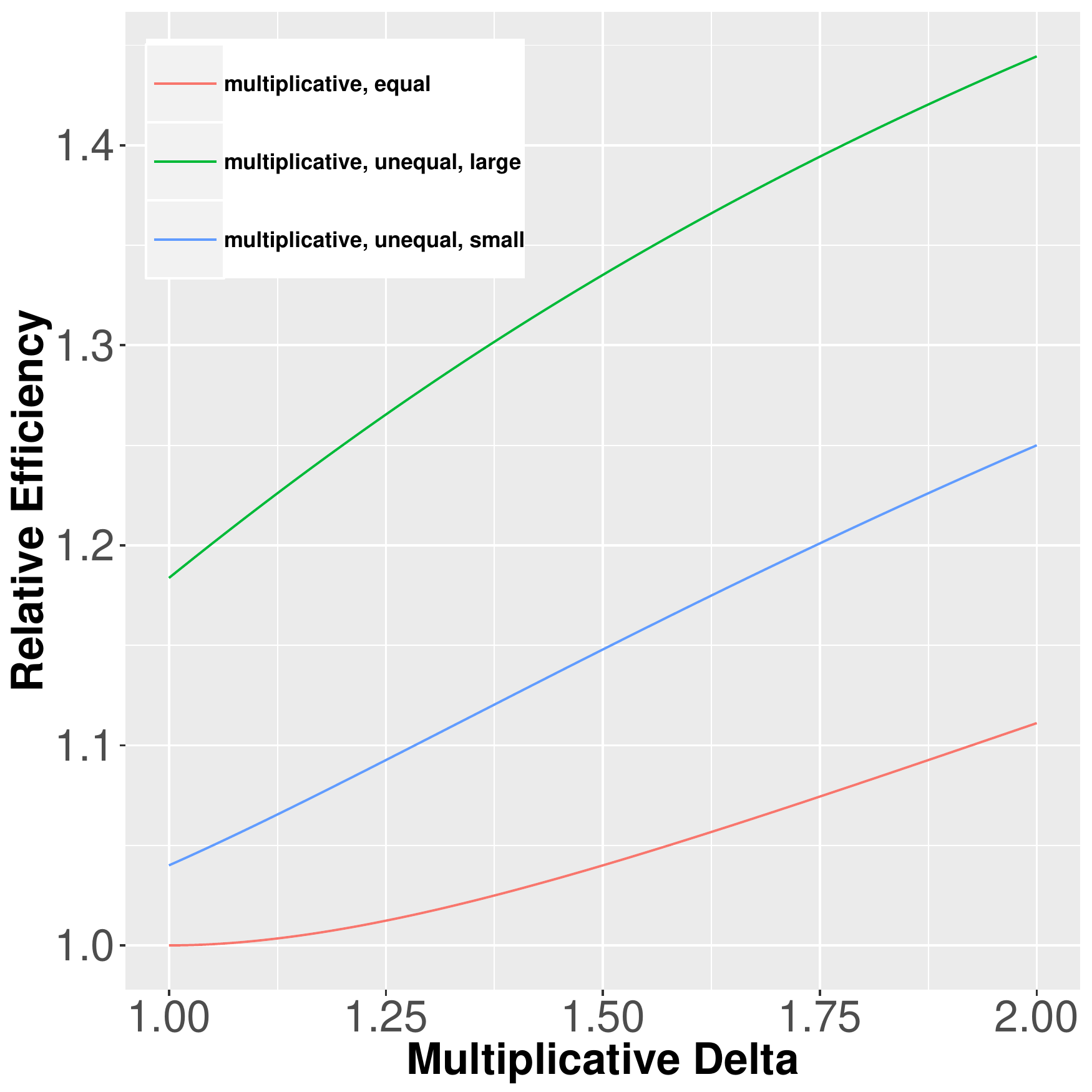}
\caption{
The asymptotic relative efficiency of optimal allocation relative to equal allocation in non-inferiority trial with multiplicative $\Delta$ and equal, unequal  variance (small difference in variance), ($\sigma_C$ = 20, $\sigma_T$ = 30) and unequal variance (large difference in variance) ($\sigma_C$ = 40, $\sigma_T$ = 100), with different $\Delta$'s.
}
\label{Fig:effdelta_multiplicative}
\end{figure}

\begin{figure}
\centering
\includegraphics[height=20pc,width=100mm]{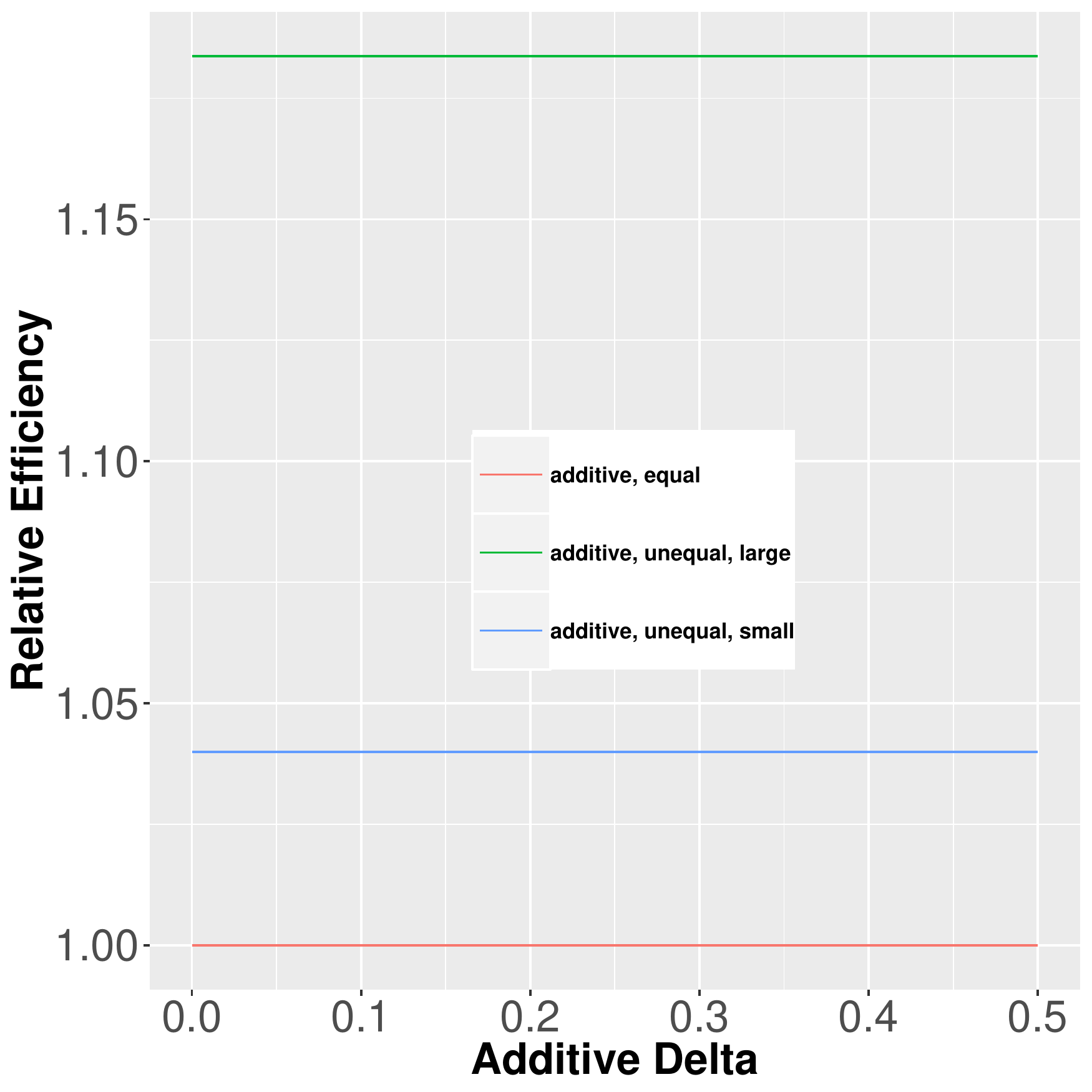}
\caption{
The asymptotic relative efficiency of optimal allocation relative to equal allocation in non-inferiority trial with additive $\Delta$ and equal, unequal  variance (small difference in variance), ($\sigma_C$ = 20, $\sigma_T$ = 30) and unequal variance (large difference in variance) ($\sigma_C$ = 40, $\sigma_T$ = 100), with different $\Delta$'s.
}
\label{Fig:effdelta_additive}
\end{figure}

Figure \ref{Fig:effdelta_multiplicative}, \ref{Fig:effdelta_additive} shows the relative efficiency of optimal allocation relative to equal allocation in non-inferiority trial with multiplicatice and additive hypothesis. 
Each line represents different non-inferiority trial parameters (either equal or unequal variance with large or small difference in variance). 
The blue and green line shows how differences with variance can affect the relative efficiency compared to a balanced allocation. 
For additive $\Delta$, the line is shown for $0<\Delta \leq 0.5$ and for multiplicative $\Delta$, the line is shown for $1 < \Delta \leq 2$.

\bibliography{wileyNJD-AMA}%

\end{document}